\documentclass[sigconf]{acmart}
\acmConference[EASE 2026]{The 30th International Conference on Evaluation and Assessment in Software Engineering}{Tue 9 - Fri 12 June 2026}{Glasgow, United Kingdom}
\renewcommand\footnotetextcopyrightpermission[1]{} 
\usepackage{subcaption}
\usepackage{caption}
\usepackage{array}
\usepackage{multirow}
\usepackage{natbib}
\usepackage{pifont}
\usepackage{placeins}
\usepackage{tabularx}
\usepackage{soul}
\usepackage{amsmath}
\usepackage{mathtools}

\definecolor{mygreen}{rgb}{0,0.6,0}
\definecolor{myred}{rgb}{0.8,0,0}

\begin{document}

\title{An Empirical Analysis of Mobile Energy Consumption Across User Configurations} 

\author{Wellington Oliveira}
\affiliation{%
 \institution{
  University of Lisbon}
 }
\email{woliveira@fc.ul.pt}

\begin{abstract}

Mobile devices have become ubiquitous tools for communication, entertainment, and productivity, yet battery autonomy remains a constraint. While energy-saving tips exist, they are often generic, anecdotal, or focused on software development rather than end-user behavior, leaving users to rely on grey literature or tacit knowledge to optimize their device energy consumption, lacking the academic rigor to ensure their effectiveness. This research aims to bridge the gap between technical energy analysis and practical user application by quantifying the energy consumption of different user-controlled parameters.
Employing an automated monitoring framework, a series of user interface tests that simulate realistic usage patterns across popular applications (i.e., WhatsApp, Instagram, TikTok, and YouTube) was conducted. The objective is to have a systematic evaluation of the energy impact of user-controllable factors, including device settings, such as screen brightness, refresh rate, connectivity status, interface themes, and battery-saving profiles, combined with more app-specific variables (e.g., video resolution and message size). By analyzing over 12,000 data points, this paper quantifies the real-world impact of common settings, revealing the trade-offs between user experience and device autonomy.

\end{abstract}


\maketitle
\section{Introduction} 
Smartphones have evolved into the central hub for modern daily life, seamlessly supporting everything from high-definition entertainment to essential productivity, communication, and navigation. However, as hardware capabilities and application complexity continue to expand, device energy demands have also increased. Although battery technology and software constructs have seen incremental improvements, battery autonomy remains a hardware bottleneck and a persistent source of frustration for users~\cite{Javed:2017,Almasri:2024}. Therefore, understanding how everyday usage patterns influence energy consumption is needed to mitigate this constraint.

Despite the universal desire to extend battery life~\cite{zaragoza2025understanding}, end-users are left to rely on generic, anecdotal, or scientifically unverified advice. While there is a body of academic literature addressing mobile energy consumption, it is overwhelmingly targeted at software developers, focusing on code-level optimizations~\cite{oliveira2023, flores2024enhancing,schuler2024systematic}. This creates a challenge: technical research rarely translates into practical, actionable guidance for the user. They end up struggling to determine which specific settings or application behaviors minimize energy drain compared to those that offer negligible benefits.

Bridging this gap requires a shift from theoretical power models to empirical, on-device measurements, reflecting realistic human-device interaction. Instead of profiling isolated components or static device states, energy consumption must be quantified while users interact with applications. By automating these interactions and monitoring energy draw directly on the smartphone, it becomes possible to isolate and measure the precise impact of user-controlled variables (e.g., screen brightness, interface themes, display refresh rates, and media resolution) under realistic conditions.

The primary objective of this research is to demystify mobile battery drain and develop empirically validated, user-centric recommendations that extend device autonomy. To achieve this, an automated monitoring framework was deployed using \texttt{ebserver}~\cite{oliveira2023ebserver} to orchestrate user interface tests across globally popular applications, specifically WhatsApp, Instagram, TikTok, and YouTube, combined with a tool functionality, the Flaslight. By systematically combining both general device settings and app-specific configurations, 879 distinct experiments were conducted. The experiments resulted in a robust dataset of over 12,000 data points. Our subsequent statistical analysis challenges several common battery-saving assumptions, revealing, for instance, that Dark Theme resulted in marginal savings compared to the impact of screen brightness, and that reducing video resolution offers negligible battery benefits.

To systematically explore these dynamics, this study addresses the following research questions:
\begin{itemize}
\item \textbf{RQ1}: How significantly do general user settings impact the energy consumption of mobile applications?
\item \textbf{RQ2}: What is the real-world energetic cost of application-specific configurations and actions?
\end{itemize}
The remainder of this paper is organized as follows: Section~\ref{sec:related} presents the related work. Section~\ref{sec:meth} details the experimental methodology. Section~\ref{sec:results} presents the data analysis and answers to the research questions. Section~\ref{sec:threats} acknowledges the limitations and threats to validity of this study. Finally, Section~\ref{sec:conclusions} concludes the paper and outlines directions for future work.

\section{Related Work}\label{sec:related}
The literature on mobile energy consumption can be divided into three main areas: technical characterization, developer-side optimization, and user-centric analysis.

\subsection{Technical Characterization}
Research in this area focuses on component-level power modeling and typically relies on hardware-based measurement. Tools such as the Monsoon Power Monitor~\cite{monsoonHVPM} provide high-precision readings directly from the power line. Studies like AndroWatts~\cite{guegain2025androwatts} and large-scale OLED profiling~\cite{dash2021much} utilize these methods to characterize device component behavior. While highly accurate, hardware setups are complex, expensive, and often require physical device alterations that prevent realistic, everyday usage simulations.

\subsection{Developer-Side Optimization}
To overcome hardware limitations and scale testing, researchers use software-based methods relying on Android's internal subsystems (e.g., \texttt{batterystats}\footnote{\url{https://developer.android.com/topic/performance/power/setup-battery-historian\#gather-data}} or the BatteryManager API\footnote{\url{https://developer.android.com/reference/android/os/BatteryManager}}) to estimate consumption based on CPU, screen, and network metrics. Developer-focused tools like GreenSource\cite{rua2019greensource} and E-MANaFA~\cite{rua2022manafa} leverage these software metrics to evaluate and optimize application source code. However, these works prioritize developer implementation choices rather than the user-configurable settings.

\subsection{User-Centric Analysis}
Recent literature has shifted toward understanding the end-user's role in energy drain. Ferreira et al.\cite{ferreira2011understanding} analyzed charging habits, while systems like Serenus\cite{lee2024serenus} predict real-time energy usage to reduce user battery anxiety and inform users about the actual energy consumed during their usage. Nevertheless, a gap remains between user intent and capability; Zaragoza et al.~\cite{zaragoza2025understanding} highlighted a critical awareness-action gap, showing that while users value energy efficiency, they lack the knowledge to optimize software usage effectively without specific feedback.

To provide actionable advice, recommendations must be based on usage rather than technical metrics. Souza et al.\cite{souza2023optimizing} proposed a system for settings like brightness, relying on static states (e.g., screen on or off) rather than dynamic application workflows. 

Aiming to help users optimize their battery usage, the present work integrates user interface automation tools\footnote{\url{https://developer.android.com/training/testing/other-components/ui-automator}} to simulate realistic user behaviors like scrolling, typing, and video playback. By wrapping these tests in an extended version of \texttt{ebserver}~\cite{oliveira2023ebserver}, it is possible to control device parameters and quantify energy drain under interactive conditions that accurately reflect daily user habits.

\section{Methodology} \label{sec:meth}

This section details the system architecture and the automated workflow designed to execute the experimental scenarios across the selected mobile applications.

\subsection{Architecture}
\begin{figure*}[h!]
\centering
\includegraphics[trim={0cm 4.2cm 0cm 4.2cm}, clip, width=0.62\textwidth]{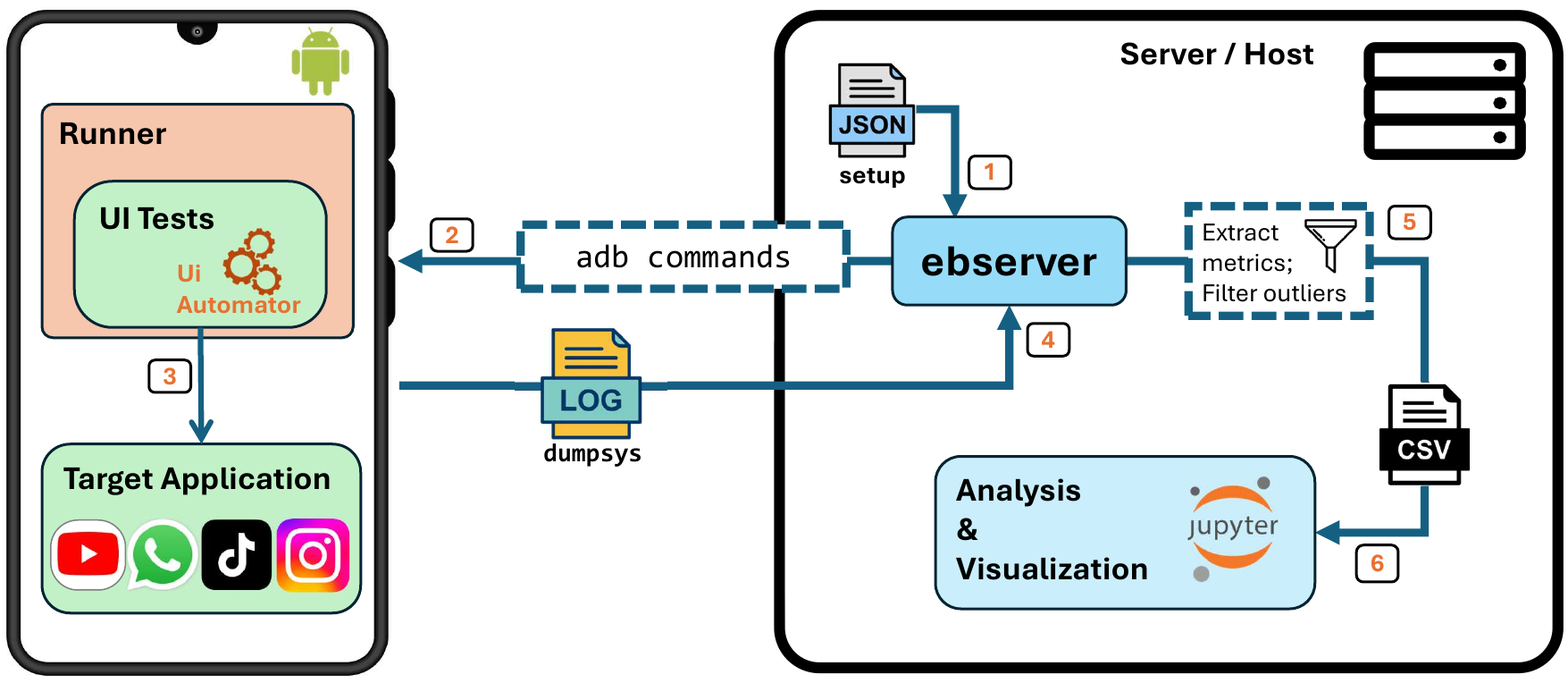}
\caption{System Architecture}
\label{fig:Architecture}
\end{figure*}

Figure \ref{fig:Architecture} illustrates the system architecture and the coordinated workflow of the experiments.
The host machine is managed by \texttt{ebserver}, which parses the \texttt{setup.json} file (\textcolor{orange}{1}). This file contains all the necessary instructions, including the sequence of tests to run and the specific parameter combinations for each execution. The configuration is highly extensible, allowing any number of general and app-specific variables to be defined for the test matrix.

Beyond configuration, \texttt{ebserver} controls the entire execution workflow. It sets the communication via the Android Debug Bridge (\texttt{adb}) (\textcolor{orange}{2}) to send commands to connect the device, install target applications, and trigger the experiments. It programmatically adjusts device-level settings, such as brightness, refresh rate, and interface theme, to match the test configuration. Before starting a run, all data is wiped from the device's \texttt{batterystats} history. This eliminates any residual from previous executions, ensuring a clean baseline. 

On the device side, a \texttt{Runner} application acts as a wrapper to orchestrate the UI testing and controlled energy measurements. Within this application, app-specific UI tests are implemented using the UI Automator. Each test is tailor-made to simulate realistic user scenarios. To call these tests, \texttt{adb} parameters were used to specify the exact general and app-specific variables (\textcolor{orange}{3}). Once a UI test completes, \texttt{ebserver} stops the monitoring process and proceeds to data collection before iterating to the next configuration.

The data collection phase relies on \texttt{adb dumpsys batterystats} to generate a detailed log file for each execution. These logs contain a breakdown of resource utilization (CPU, GPU, memory) and fine-grained energy consumption metrics. This includes both the overall energy drained by hardware components (e.g., screen, Wi-Fi) and the specific energy footprint of the target application. The logs are automatically retrieved, organized, and categorized on the host (\textcolor{orange}{4}).

The raw data moves to a pre-processing pipeline (\textcolor{orange}{5}). Because the necessary metrics are scattered throughout the \texttt{dumpsys} output, regular expressions are used to parse the logs and extract the targeted energy values. Given that minor background OS processes or external factors can occasionally introduce noise into mobile environments, statistical outliers are filtered out from the dataset. The output of this pipeline is a structured \texttt{CSV} file dataset.

In the final step (\textcolor{orange}{6}), the dataset is analyzed to quantify the precise energy impact of each evaluated variable. This analysis, done using Jupyter Notebook, compares the energy footprints across different configuration combinations, enabling the extraction of statistically validated insights regarding user behavior and battery drain.

\subsection{Target Applications and Use Cases}

The selected applications were chosen as representative archetypes of the most common smartphone user behaviors, prioritizing global popularity and high usage rates. For instance, WhatsApp was selected to represent messaging apps; TikTok and Instagram represent media-heavy social networks, characterized by continuous scrolling; Finally, YouTube was selected as the benchmark for video streaming. Table~\ref{tab:AppsUseCases} details the target applications, including their experimental aliases, popularity metrics, and the specific automated UI interactions designed for this study.

\begin{table*}[ht]
\centering
\small
\caption{Applications and their characteristics}
\label{tab:AppsUseCases}
\begin{tabularx}{\textwidth}{l l l l l >{\raggedleft\arraybackslash}X}
\toprule
\textbf{Application} & \textbf{Short Form} & \textbf{Category} & \textbf {Reviews} & \textbf{Downloads} &  \textbf{UI Actions} \\
\midrule
\textbf{YouTube} 
    & YT & Video & 150M+ & 10B+
    & start video; select resolution; skip to two-thirds of video \\
\textbf{WhatsApp} 
    & WPP & Messaging & 191M+ & 5B+
    & navigate to chat; type message; send message \\
\textbf{Instagram} 
    & IG & Social Media & 148M+ & 5B+ 
    & scroll posts; wait; like posts \\
\textbf{TikTok} 
    & TT & Social Media & 69M+ & 1B+
    & scroll posts; wait; pause  \\
\textbf{Flashlight} 
    & LIGHT & Tool & N/A & N/A
    & start flashlight  \\
\bottomrule
\end{tabularx}
\end{table*}

\subsection{Experimental Control Variables}

To isolate and measure the energy footprint of specific device operations, experiments were designed using strict control variables. These variables are divided into two categories: \textit{general variables} and \textit{app-specific variables}, as summarized in Table~\ref{tab:controlled_variables}.

\vspace{0.2cm}
\noindent
\textbf{General Variables:}~These are system-wide settings that directly impact device hardware, such as screen brightness, refresh rate, interface theme, and connectivity (Airplane Mode). These variables are independent of any specific application. Our framework allows for the selective exclusion of these variables when irrelevant (e.g., screen refresh rate is omitted when testing the Flashlight). Notably, the \textsc{Power Saving} mode acts as an overriding variable. While the system claims it applies several restrictions, only the following actually affected our test environment: (i) limiting CPU speed to 70\%, (ii) disabling Always-On Display, (iii) forcing Dark  Mode\footnote{Different effects from Dark theme; it greatly reduced the OS graphical elements.}. 

\vspace{0.2cm}
\noindent
\textbf{App-Specific Variables:}~These parameters present how the user interacts with a specific application, controlling the exact workload on the app. A subset of variables that represent the most common real-world usage scenarios was selected for the experiments.

\begin{table}[ht]
\small
\centering
\caption{General and app-specific experimental variables}
\label{tab:controlled_variables}
\begin{tabular}{lll}
\toprule
\textbf{Category / App} & \textbf{Variable} & \textbf{Settings} \\
\midrule
\multirow{5}{*}{\textbf{General}} 
 & Theme & Light / Dark \\
 & Screen brightness (\%) & 0 / 50 / 100 \\
 & Refresh rate (Hz) & 30 / 60 / 120 \\
 & Airplane mode & On / Off \\
 & Power saving mode & On / Off \\
\midrule
\textbf{Flashlight} & Duration (s) & 15 / 30 / 60 \\
\textbf{Flashlight} & Intensity (\%) & 0 / 10 / 25 / 50 / 100 \\
\textbf{YouTube} & Video Resolution & 240p / 720p / 1440p \\
\textbf{YouTube} & Duration (s) & 15 / 30 \\
\textbf{Instagram} & Duration (s) & 15 / 30 \\
\textbf{TikTok} & Duration (s) & 15 / 30 \\
\textbf{WhatsApp} & Message Size (chars) & 100 / 200 \\
\bottomrule
\end{tabular}
\end{table}

\subsection{Execution Scenarios}

To simulate realistic behavior, distinct use cases were scripted for each application. All general variables were applied across the application tests, with the exception of the Flashlight experiment, which only used duration and intensity due to its simplicity.

\begin{itemize}
    \item \textbf{WhatsApp:} The app is opened, navigated to a specific chat, and a message is sent. The app-specific variable is the message size (100 or 200 characters), utilizing predefined text snippets hardcoded into the automation script.
    \item \textbf{TikTok \& Instagram:} Both scripts simulate continuous scrolling. The TikTok script scrolls for a set duration, followed by a pause/play action midway. The Instagram script is similar but replaces the pause with a double-click (``like'') on the current post before resuming the scrolling loop.
    \item \textbf{YouTube:} The script opens a specific video, sets the target resolution, skips to two-thirds of the video's duration, and plays for the set time. The skip is needed to bypass different resolution segments that may have been pre-buffered. The video\footnote{\url{https://youtu.be/njX2bu-_Vw4}} was chosen because its leverages the properties of OLED screens. Keeping the same video across all iterations eliminates confounding variables such as video compression quality, color palette, and audio bitrates.
\end{itemize}

\subsection{Dataset Generation}

To evaluate the energy impact across these scenarios, the total number of experimental configurations ($E$) was calculated as the Cartesian product of the general variables ($S$) and the app-specific variables ($C_a$). $|S|$ is defined by the formula:
\begin{equation}
|S| = |T| \times |B| \times |R| \times |P| \times |A|
\end{equation}
Where $T$ represents interface themes (2), $B$ screen brightness levels (3), $R$ refresh rates (3), $P$ battery saving profile (2), and $A$ airplane mode (2). Therefore $|S| = 2 \times 3 \times 3 \times 2 \times 2 = 72$.

Given a set of target applications $A$, with $C_a$ representing the number of app-specific configurations tested for an application $a \in A$. The total number of unique experimental configurations ($E$) is defined as:
\begin{equation}\label{form:summation}
E = \left( |S| \times \sum_{a \in A} C_a \right) + E_{flashlight}
\end{equation}

The app-specific parameters were structured as follows: $C_{IG} = C_{TT} = 2$ (\textsc{Duration}), $C_{WPP} = 2$ (\textsc{Message size}), and $C_{YT} = 6$ (3 \textsc{Video Resolution} $\times$ 2 \textsc{Duration}). The Flashlight experiments ($E_{flashlight}$) were executed independently of the general variables, resulting in $15$ unique states (5 \textsc{Intensity} $\times$ 3 \textsc{Duration}). Replacing the these values: $E = (72 \times 12) + 15 = 879$ unique configurations.

To ensure statistical significance and maintain a manageable execution timeframe, $I = 15$ iterations were run for each configuration, resulting in an initial dataset of 13,184 data points. After applying statistical filters, 4.0\% (536) of the points were flagged as outliers and excluded, leaving 12,649 data points for the final analysis.

All experiments were conducted on a Samsung Galaxy S23 Ultra smartphone, equipped with a Snapdragon 8 Gen 2 processor, a 5000 mAh battery, and a 6.8-inch Dynamic AMOLED display capable of a 120Hz refresh rate. Raw energy consumption values, originally extracted from the \texttt{batterystats} in mAh, were converted to Joules using the instantaneous voltage present in the system logs~\cite{sharma2025critical}.

\section{Experimental Results}\label{sec:results}

Since the collected data did not follow a normal distribution, the Mann-Whitney U test was employed to determine statistical significance. Consequently, only statistically significant results ($p < 0.05$) are presented individually. The complete dataset containing all experimental results is available in the replication repository~\cite{repository}. The reported values are the average for that specific combination.

To effectively visualize the data variance, all plots utilize a logarithmic scale. In these representations, positive values indicate an increase in energy consumption, whereas negative values represent a decrease. For comparative analysis, energy consumption was consistently measured in Watts (Joules per second). The exceptions are the variables \textsc{Duration} and \textsc{Message Size}; since average power across different time intervals does not reflect the accumulated battery drain, total energy (Joules) was used. Finally, some graphs include an \textsc{Overall} bar, which represents the aggregate expected impact of a specific variable across all evaluated applications. This metric is composed of the average of all other averages in order to mitigate the bias from the different numbers of experimental data from each application. To ensure a realistic global representation, this aggregation incorporates all collected data points.

\begin{figure}[htbp]
    \centering
    \captionsetup[subfigure]{skip=0pt} 

    \begin{subfigure}[b]{0.43\textwidth}
        \centering
        \includegraphics[trim={0cm 0cm 0cm 0cm}, clip, width=\linewidth]{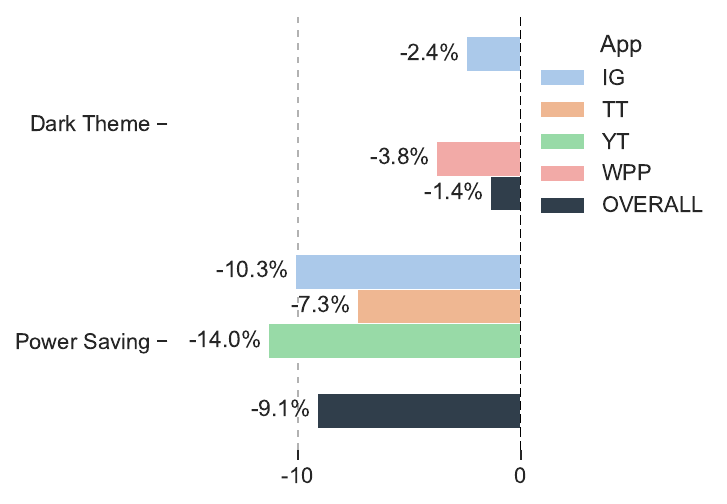}
        \caption{\textsc{Dark Theme} and \textsc{Battery Saving}}
        \label{fig:binary_res}
    \end{subfigure}
    \hfill 
    \begin{subfigure}[b]{0.45\textwidth}
        \centering
        \includegraphics[trim={2cm 0cm 4cm 0cm}, clip, width=\linewidth]{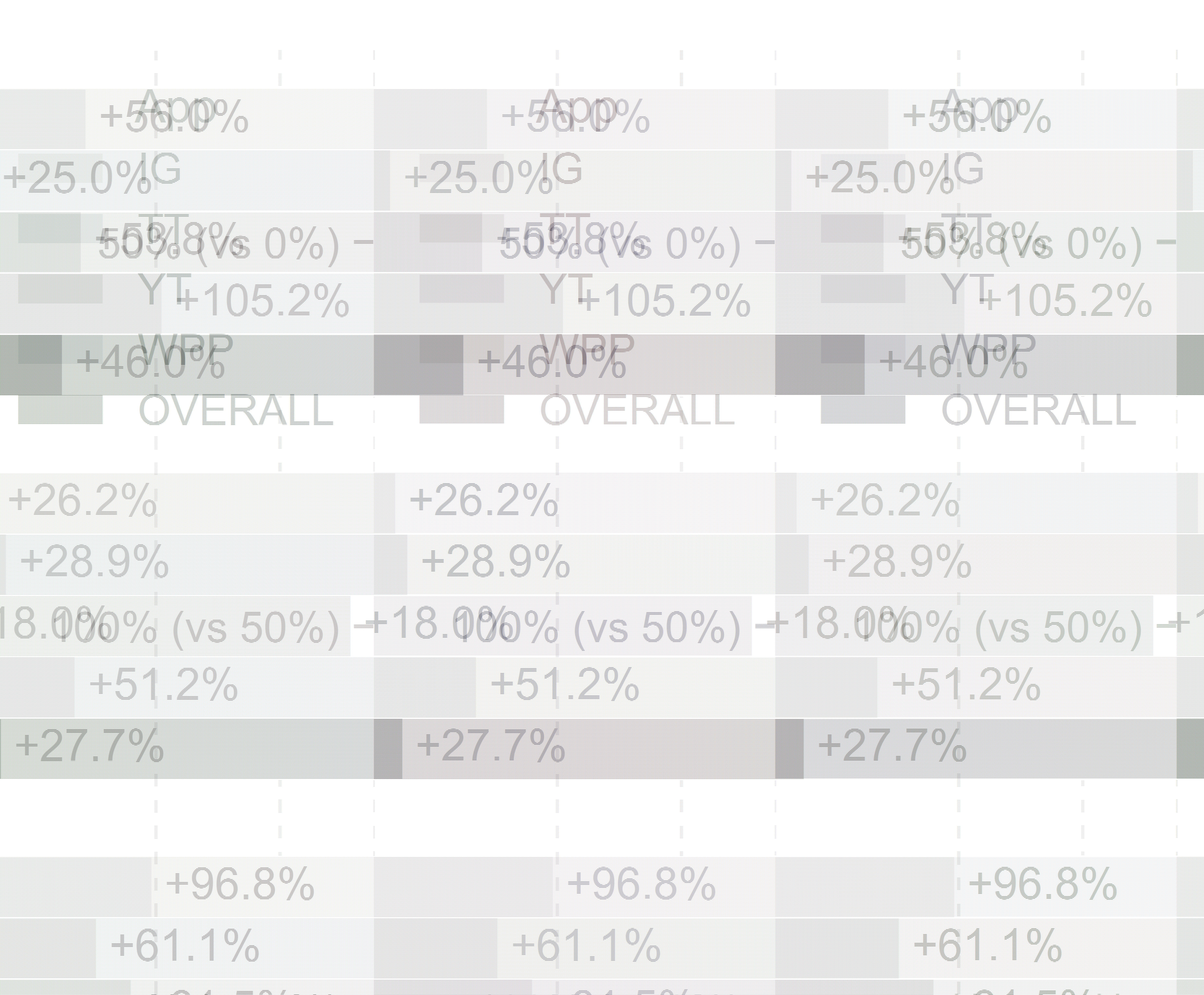}
        \caption{\textsc{Brightness}}
        \label{fig:bright_res}
    \end{subfigure}
    
    \par\bigskip 
    
    \begin{subfigure}[b]{0.45\textwidth}
        \centering
        \hbox{\hspace{-0.4em}\includegraphics[trim={2cm 0cm 4cm 0cm}, clip, width=\linewidth]{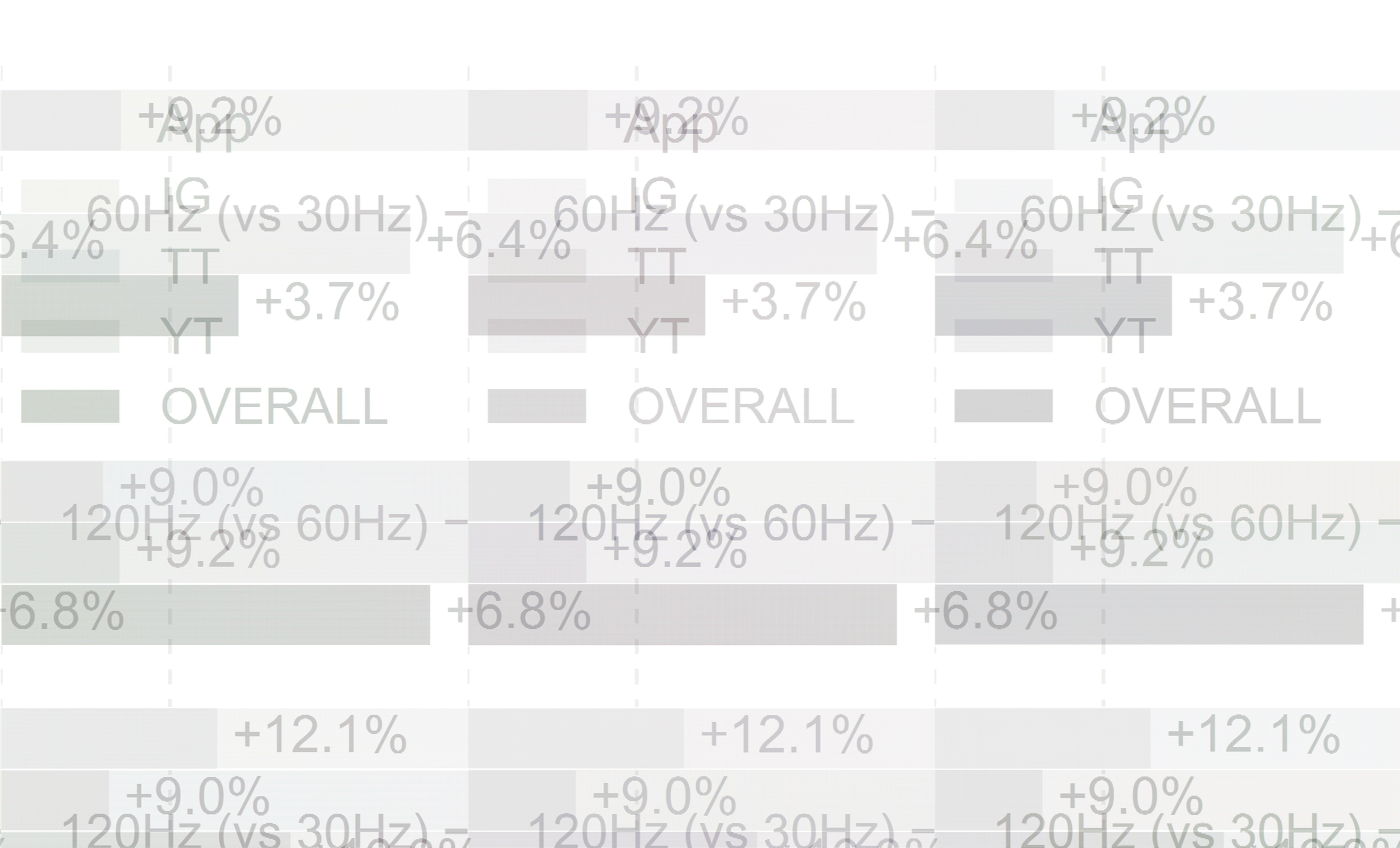}}
        \caption{\textsc{Refresh Rate}}
        \label{fig:rate_res}
    \end{subfigure}
    
    \caption{General Variables Power Impact}
    \label{fig:general_res}
\end{figure}

\vspace{0.2cm}
\noindent
\textbf{General Variables:}~Figure~\ref{fig:general_res} summarises the energy impact of the general variables evaluated. \textsc{Airplane mode} is not present because it resulted in no statistically significant effect.

Figure~\ref{fig:binary_res} reports the binary general variables, \textsc{Theme} and \textsc{Power Saving}. Enabling the \textsc{Dark} theme reduced energy consumption for two applications: Instagram by 2.4\% and WhatsApp by 3.8\% relative to \textsc{Light} theme, with an overall reduction of 1.4\%. In contrast, enabling \textsc{Power Saving} (\textsc{On}) reduced energy consumption for all applications except WhatsApp. YouTube exhibited the largest reduction (14.0\%), whereas TikTok exhibited the smallest (7.3\%). Averaged across applications, enabling \textsc{Power Saving} reduced energy consumption by 9.1\%.

\textsc{Brightness} produced the largest and most consistent effect across applications (Figure~\ref{fig:bright_res}), which compares energy overhead when running the same workload at different brightness levels. All applications showed increased consumption with higher brightness. The effect was particularly pronounced for Instagram and WhatsApp when moving from 0\% to 100\% brightness (96.8\% and 210.2\% increases, respectively). Overall, increasing brightness from 0\% to 100\% increased energy consumption by 86.5\% on average.

Finally, Figure~\ref{fig:rate_res} reports the effect of \textsc{Refresh Rate}. Increasing the refresh rate increased energy consumption for all applications except WhatsApp, which was the only workload that did not involve videos. TikTok was unaffected when increasing from 30Hz to 60Hz, whereas the remaining applications showed modest increases (e.g., Instagram increased by 9.2\%). Using the maximum available refresh rate increased energy consumption by up to 16.3\% for YouTube. Overall, increasing the refresh rate from 30Hz to 120Hz increased energy consumption by 10.8\% on average.

\begin{figure}[htbp]
    \centering
    \captionsetup[subfigure]{skip=0pt} 

    \begin{subfigure}[b]{0.47\textwidth}
        \centering
        \hbox{\hspace{-0.3em} \includegraphics[trim={0cm 0cm 0cm 0cm}, clip, width=\linewidth]{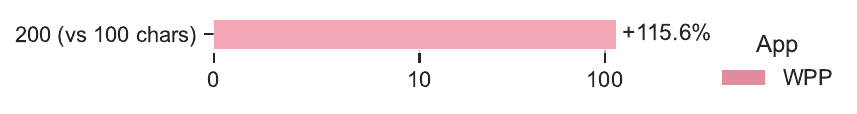}}
        \caption{\textsc{Message size} in Whatsapp}
        \label{fig:wpp_res}
    \end{subfigure}
    \hfill 
    \begin{subfigure}[b]{0.45\textwidth}
        \centering
        \hbox{\hspace{-0.3em} \includegraphics[trim={0cm 0cm 0cm 0cm}, clip, width=\linewidth]{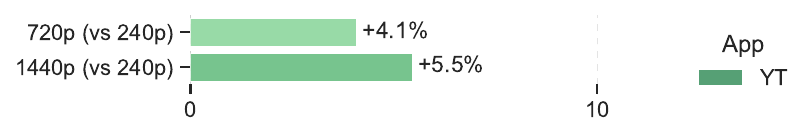}}
        \caption{\textsc{Video Resolution} in YouTube}
        \label{fig:yt_res}
    \end{subfigure}
    
    \par\bigskip 
    
    \begin{subfigure}[b]{0.48\textwidth}
        \centering
        \hbox{\hspace{-0.2em} \includegraphics[trim={0cm 0cm 0cm 0cm}, clip, width=\linewidth]{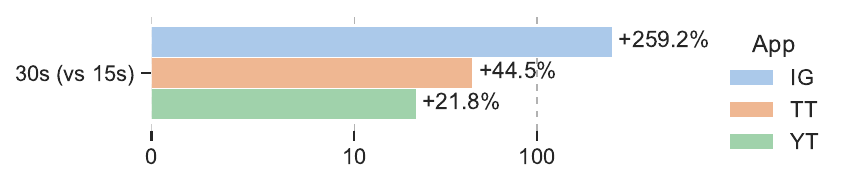}}
        \caption{\textsc{Duration} in Instagram, TikTok and YouTube}
        \label{fig:dur_res}
    \end{subfigure}
    \hfill
    \begin{subfigure}[b]{0.45\textwidth}
        \centering
        \hbox{\hspace{-0.9em} \includegraphics[trim={0cm 0cm 0cm 0cm}, clip, width=\linewidth]{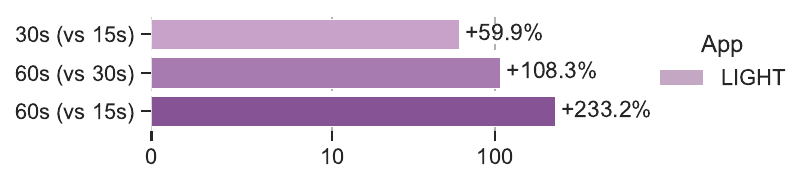}}
        \caption{\textsc{Duration} in Flashlight}
        \label{fig:light_res}
    \end{subfigure}
    
    \caption{App-Specific Variables Impact}
    \label{fig:specific_res}
\end{figure}

\vspace{0.2cm}
\noindent
\textbf{App-Specific Variables:}~Figure~\ref{fig:specific_res} summarises the energy impact of the app-specific variables. \textsc{Intensity} for the Flashlight is excluded because it did not produce statistically significant differences. This suggests that, within the \textsc{Duration} used, users do not need need to consider energy consumption when selecting flashlight intensity.

All other app-specific variables resulted in statistically significant effects: \textsc{Message Size} for WhatsApp (Figure~\ref{fig:wpp_res}), \textsc{Video Resolution} for YouTube (Figure~\ref{fig:yt_res}), and \textsc{Duration} for (i) YouTube, Instagram, and TikTok, and (ii) the Flashlight (Figures~\ref{fig:dur_res} and~\ref{fig:light_res}, respectively).

For WhatsApp, doubling the number of characters increased energy consumption by 115.6\%. For YouTube, selecting the highest tested video resolution increased energy consumption by 5.5\%; however, the difference between 720p and 1440p was not statistically significant. The \textsc{Duration} results are presented in two figures: for most applications, the differences between 15s and 30s were compared; for the Flashlight, 60s were also included. Instagram showed the largest effect for \textsc{Duration}, increasing by 259\% from 15s to 30s. TikTok and YouTube exhibited more moderate increases (44.5\% and 21.8\%, respectively). For the Flashlight, the effect was approximately linear: relative to 15s, consumption increased by 59.9\% at 30s and by 233.2\% at 60s.

\subsection{Discussion}
The results in this paper provide practical guidance for users who want to reduce smartphone energy consumption through settings and in-app choices. For the general variables, \textsc{Dark Theme} only reduced energy for applications whose interface is heavily influenced by the theme, and even then, the gains were small. Therefore, \textsc{Dark Theme} should not be a primary energy-saving strategy, particularly if it leads users to increase \textsc{Brightness} to maintain readability.

\textsc{Brightness} is the dominant driver of energy consumption across applications. Although 0\% brightness may be impractical, already moving from 0\% to 50\% incurs a substantial battery cost. Therefore, brightness should be kept as low as usability allows. Enabling adaptive brightness (i.e., automatically adjusting brightness to ambient light) might be an effective strategy to reduce energy consumption without sacrificing usability.

\textsc{Refresh Rate} increased energy consumption in most scenarios, with the largest overhead at 120Hz. For many users, 60Hz represents a good compromise: it captures most perceptual benefits such as motion smoothness and reduced flickering while avoiding the diminishing returns at higher refresh rates~\cite{rao:2022}.

\vspace*{0.2cm}
\noindent
\hspace{0.01em}
\fcolorbox{black!75!black}{blue!3!white}{%
\parbox{0.96\columnwidth}{%
\textbf{RQ1:} Across general variables, \textsc{Airplane mode} did not reduce energy in our workloads, and \textsc{Dark Theme} produced only marginal savings (1.4\% overall) while potentially reducing readability and encouraging higher brightness. In contrast, \textsc{Brightness} is the main lever: 50\% setting raises energy consumption by 46\% overall, so it should be minimized within comfort. For \textsc{Refresh Rate}, 60Hz is a practical default, increasing power by only 3.7\% while preserving a smooth experience.
}}
\vspace{0.2cm}

For app-specific variables, \textsc{Message Size} in WhatsApp increased energy consumption approximately linearly, as expected. On the other hand, YouTube energy consumption appears to be driven primarily by video processing~\cite{Kazantsev:2021}, with \textsc{Video Resolution} having only a modest effect. Consequently, reducing resolution should be considered a last-resort energy-saving measure.

The effect of \textsc{Duration} is more nuanced. On YouTube, doubling the time did not double energy consumption, consistent with a large fixed cost associated with video processing and system overheads. Instagram, on the other hand, showed a larger increase, which may relate to additional background activity and the presence of advertisements, known contributors to energy consumption~\cite{Pathak2012Eprof}. TikTok, despite similar usage patterns as Instagram, did not exhibit the same spike, suggesting that application design and content delivery pipelines might affect the consumption more than time alone.

\vspace*{0.2cm}
\noindent
\hspace{0.2em}
\fcolorbox{black!75!black}{blue!3!white}{%
\parbox{0.93\columnwidth}{%
\textbf{RQ2:} For app-specific factors, \textsc{Video Resolution} had a negligible effect on energy and is an unlikely knob for saving battery. \textsc{Message Size} scaled approximately linearly with energy. By contrast, \textsc{Duration} was not reliably linear: costs are shaped more by workload characteristics such as video processing and by app behaviour, which can outweigh execution time alone.
}}

\section{Limitations}\label{sec:threats}

While this study provides a comprehensive, user-centric analysis of mobile energy consumption in real-world scenarios, certain limitations must be acknowledged.

Foremost, our data acquisition relies on Android's software-based power model rather than external hardware power monitors. This methodology is the established standard for scalable, on-device energy profiling and has been extensively validated in recent literature~\cite{huber:2022, oliveira2023}. Furthermore, previous studies have demonstrated that \texttt{batterystats} yields an accuracy level that is comparable to physical measurement instruments~\cite{monteiro2023analysis}, making it a reliable tool for comparative energy evaluations.

A second limitation concerns the generalizability of the results across diverse mobile devices. The experiments were conducted on a single device (Samsung Galaxy S23 Ultra). While our findings might be accurate for modern flagship architectures, replicating this framework across a wider spectrum of budget and legacy devices would enhance its generalizability.

Finally, while the automated test scenarios were designed to simulate realistic user behaviors, they abstract the unpredictability of human interaction. Real-world usage involves multitasking, changing network signals, and sensor interference that a controlled experimental setup cannot encapsulate. Nonetheless, isolating these specific variables is needed to quantify their energetic footprint and to create reliable, evidence-based recommendations for end-users.

\section{Conclusions} \label{sec:conclusions}

This study bridges the gap between technical energy metrics and practical user behavior by quantifying the battery impact of everyday settings across popular mobile applications. Our analysis of over 12,000 data points challenges common energy-saving assumptions: Dark Theme offers only marginal savings (1.4\%), whereas screen brightness is the dominant factor, increasing energy consumption by an average of 86.5\% at maximum levels. Furthermore, capping refresh rates at 60 Hz emerged as the optimal efficiency trade-off, while lowering video resolution resulted in negligible benefits, mostly due to the fixed energy costs of video processing.

Future work will expand this study by incorporating a wider variety of mobile devices, application categories, different apps from the same category (e.g., Telegram, Signal), more app-specific variables, an increase number of metrics (e.g., network-usage) and different use cases. Another promising avenue involves leveraging the generated dataset to train supervised machine learning models. Ultimately, this project aims to translate these technical findings into a practical, user-oriented guide, providing evidence-based recommendations to help users maximize their device autonomy.

\begin{acks}
This work is funded by national funds through FCT – Fundação para a Ciência e a Tecnologia under the LASIGE Research Unit, ref. UID/00408/2025, DOI \textcolor{blue}{\url{https://doi.org/10.54499/UID/00408/2025}} MOSA - Mobile Optimization on Security through Algorithmic Improvement, project under the reference 2024.17405.PEX, DOI \textcolor{blue}{\url{https://doi.org/10.54499/2024.17405.PEX}}
\end{acks}

\bibliography{sample-base}

\end{document}